\documentclass[twoside]{article}
\usepackage{fleqn,espcrc2}

\usepackage{graphicx}
\usepackage[figuresright]{rotating}

\newcommand{\AmS}{{\protect\the\textfont2
  A\kern-.1667em\lower.5ex\hbox{M}\kern-.125emS}}

\title{Higgs- and quark-inspired modifications 
of the finite-temperature properties of the 
Polyakov model}

\author{D. Antonov\address{INFN-Sezione di Pisa, 
 Universit\'a degli studi di Pisa, \\
 Dipartimento di Fisica, 
 Via Buonarroti, 2 - Ed. B - I-56127 Pisa, Italy}}

\begin{document}

\begin{abstract}
(2+1)-dimensional Georgi-Glashow model, else called the Polyakov model, is explored 
at nonzero temperatures and in the
regime when the Higgs boson is not infinitely heavy.
The finiteness of the Higgs-boson mass leads to 
the appearance of the upper bound on the 
parameter of the weak-coupling approximation,
necessary to maintain the stochasticity of the Higgs vacuum.
The modification of the finite-temperature behavior of the 
model emerging due to the introduction of massless
quarks is also discussed.
\end{abstract}

\maketitle

\section{Introduction. The model}
Since the second half of the seventies~\cite{1},
(2+1)-dimensional Georgi-Glashow model, oftenly also called the Polyakov model, is known as
an example of the theory allowing for an analytic description
of confinement. However, confinement in the Polyakov model is typically discussed
in the limit of infinitely large Higgs-boson mass, when the
model is reduced to compact QED. In the present talk, we shall discuss various  
modifications of the finite-temperature properties of the Polyakov model stemming 
from the finiteness of this mass.

The Euclidean action of the 
model reads

$$
S=\int d^3x\left[\frac{1}{4g^2}\left(F_{\mu\nu}^a\right)^2+
\frac12\left(D_\mu\Phi^a\right)^2+\right.$$

\begin{equation}
\label{GG}
+\left.\frac{\lambda}{4}\left(
\left(\Phi^a\right)^2-\eta^2\right)^2\right].
\end{equation}
Here, the Higgs field $\Phi^a$ transforms by the adjoint representation, 
$D_\mu\Phi^a\equiv\partial_\mu\Phi^a+\varepsilon^{abc}A_\mu^b
\Phi^c$. Next, $\lambda$ is the Higgs coupling constant of dimensionality [mass], 
$\eta$ is the Higgs v.e.v. of dimensionality $[{\rm mass}]^{1/2}$, and 
$g$ is the electric coupling constant of the same dimensionality.
At the one-loop level, the sector of the theory~(\ref{GG}) containing 
dual photons and Higgs bosons is represented by the following partition function~\cite{dietz}:

$$
{\mathcal Z}=1+$$

$$+\sum\limits_{N=1}^{\infty}\frac{\zeta^N}{N!}\left(\prod\limits_{i=1}^{N}\int d^3z_i
\sum\limits_{q_i=\pm 1}^{}\right)
\exp\left[-\frac{g_m^2}{8\pi}\times\right.$$

$$
\left.\times\sum\limits_{{a,b=1\atop a\ne b}}^{N}\left(\frac{q_aq_b}{|{\bf z}_a-{\bf z}_b|}-
\frac{{\rm e}^{-m_H|{\bf z}_a-{\bf z}_b|}}{|{\bf z}_a-{\bf z}_b|}\right)\right]\equiv
$$

\begin{equation}
\label{pf}
\equiv\int {\mathcal D}\chi{\mathcal D}\psi {\rm e}^{-S},
\end{equation}
where

$$
S=\int d^3x\left[\frac12(\partial_\mu\chi)^2+\frac12(\partial_\mu\psi)^2
+\frac{m_H^2}{2}\psi^2-\right.
$$

\begin{equation}
\label{1}
-\left.2\zeta{\rm e}^{g_m\psi}\cos(g_m\chi)\right]
\equiv\int d^3x{\mathcal L}[\chi,\psi|g_m,\zeta].
\end{equation}
The partition function~(\ref{pf}) describes the grand canonical ensemble of monopoles with the 
account for their Higgs-mediated interaction.
In Eqs.~(\ref{pf}) and~(\ref{1}), $\chi$ is the dual-photon field, and the field $\psi$ accounts for the Higgs field,
whose mass reads $m_H=\eta\sqrt{2\lambda}$. Note that from Eq.~(\ref{pf}) it is straightforward to deduce that when $m_H$
formally tends to infinity, one arrives at the conventional sine-Gordon theory of the 
dual-photon field~\cite{1} describing the compact-QED limit of the model.
Next, in the above equations, $g_m$ stands for the magnetic coupling constant 
related to the electric one as $g_mg=4\pi$, and 
the monopole fugacity $\zeta$ has the form
$\zeta=\frac{m_W^{7/2}}{g}\delta\left(\frac{\lambda}{g^2}\right)
{\rm e}^{-4\pi m_W\epsilon/g^2}$.
In this formula, $m_W=g\eta$ is the W-boson mass, 
and $\epsilon=\epsilon(\lambda/g^2)$ is a certain monotonic, slowly 
varying function, $\epsilon\ge 1$, $\epsilon(0)=1$~\cite{bps},  
$\epsilon(\infty)\simeq 1.787$~\cite{kirk}.
As far as the function $\delta$ is concerned, 
it is determined by the loop corrections. In what follows, we shall work 
in the standard weak-coupling regime $g^2\ll m_W$, which parallels the requirement 
that $\eta$ should be large enough to ensure the spontaneous symmetry breaking from 
$SU(2)$ to $U(1)$. The W-boson mass will thus play the role of the UV cutoff in the further analysis.

\section{The model at finite temperature beyond the compact-QED limit}

In the discussion of finite-temperature properties of the Polyakov model in the
present Section, we shall follow Ref.~\cite{plb}.
At finite temperature $T\equiv1/\beta$, one should supply the fields $\chi$ and $\psi$
with the periodic boundary conditions in the temporal direction, with the period equal to
$\beta$. Because of that, the lines of magnetic field emitted by a monopole cannot cross
the boundary of the one-period region and consequently, at the distances larger than $\beta$,  
should go almost parallel to this boundary, approaching it. 
Therefore, monopoles separated by such distances
interact via the 2D Coulomb potential, rather than the 3D one. Since the average distance
between monopoles in the plasma is of the order $\zeta^{-1/3}$, we see that at $T\ge{\cal O}\left(\zeta^{1/3}\right)$,
the monopole ensemble becomes two-dimensional. Owing to the fact that $\zeta$ is exponentially 
small in the weak-coupling regime under discussion, the idea of dimensional reduction is perfectly applicable
at the temperatures of the order of the critical temperature of the Berezinsky-Kosterlitz-Thouless (BKT)
phase transition~\cite{bkt} (for a review see e.g. Ref.~\cite{rev}) in the monopole plasma, 
which is equal to $g^2/2\pi$~\cite{2}~\footnote{Note that due to the $T$-dependence of the strength of the 
monopole-antimonopole interaction, which is a consequence of the dimensional reduction, the BKT 
phase transition in the Polyakov model is inverse with respect to the standard one of the 2D 
XY model. Namely, monopoles exist in the plasma phase at the temperatures below the BKT critical one
and in the molecular phase otherwise.}. Up to exponentially small corrections, this temperature is 
unaffected by the finiteness of the Higgs-boson mass. This can be seen from the expression for the 
mean squared separation in the monopole-antimonopole molecule, 

$$
\left<L^2\right>=
\frac{\int\limits_{|{\bf x}|>m_W^{-1}}^{} d^2{\bf x}|{\bf x}|^{2-\frac{8\pi T}{g^2}}J}{\int
\limits_{|{\bf x}|>m_W^{-1}}^{} 
d^2{\bf x}|{\bf x}|^{-\frac{8\pi T}{g^2}}J},$$
where $J\equiv\exp
\left[\frac{4\pi T}{g^2}K_0\left(m_H|{\bf x}|
\right)\right]$ and 
$K_0$ denotes the modified Bessel function.
Disregarding the exponential factors in the numerator and denominator of this equation, we obtain 
$\left<L^2\right>\simeq\frac{4\pi T-g^2}{2m_W^2\left(2\pi T-g^2\right)}$,
that yields the above-mentioned value of the BKT critical temperature $g^2/2\pi$. 
Besides that, it is straightforward to see that in the weak-coupling 
regime under study, the value of $\sqrt{\left<L^2\right>}$ is exponentially smaller than the characteristic distance in the 
monopole plasma, $\zeta^{-1/3}$, i.e., molecules are very small-sized with respect to that distance.

The factor $\beta$ at the 
action of the dimensionally-reduced theory, $S_{{\rm d.-r.}}=\beta\int d^2x{\mathcal L}[\chi,\psi|g_m,\zeta]$,
can be removed [and this action can be cast to the original form of eq.~(\ref{1}) with the substitution $d^3x\to d^2x$]
by the obvious rescaling: 
$S_{{\rm d.-r.}}=\int d^2x{\mathcal L}\left[\chi^{\rm new},\psi^{\rm new}|\sqrt{K},\beta\zeta\right]$.
Here, $K\equiv g_m^2T$, 
$\chi^{{\rm new}}=\sqrt{\beta}\chi$, $\psi^{{\rm new}}=\sqrt{\beta}\psi$, and in what follows
we shall denote for brevity $\chi^{{\rm new}}$ and $\psi^{{\rm new}}$ simply as $\chi$ and $\psi$, respectively.
Averaging then over the field $\psi$ with the use of the cumulant expansion we arrive at the 
following action:

$$
S_{{\rm d.-r.}}\simeq\int d^2x\left[\frac12(\nabla\chi)^2-2\xi\cos\left(g_m\sqrt{T}\chi\right)\right]-
$$

$$
-2\xi^2\int d^2xd^2y\cos\left(\sqrt{K}\chi({\bf x})\right){\mathcal K}^{(2)}({\bf x}-{\bf y})\times
$$

\begin{equation}
\label{2}
\times\cos\left(\sqrt{K}\chi({\bf y})\right).
\end{equation}
In this expression, we have disregarded all the cumulants higher 
than the quadratic one, and the limits of applicability of 
this so-called bilocal approximation will be discussed below. 
Further, in Eq.~(\ref{2}), ${\mathcal K}^{(2)}({\bf x})\equiv{\rm e}^{KD_{m_H}^{(2)}({\bf x})}-1$, where  
$D_{m_H}^{(2)}({\bf x})\equiv K_0(m_H|{\bf x}|)/2\pi$ is the 2D Yukawa propagator, and 
$\xi\equiv\beta\zeta{\rm e}^{\frac{K}{2}D_{m_H}^{(2)}(m_W^{-1})}$
denotes the monopole fugacity modified by the interaction of monopoles via the Higgs field.
Clearly, in the compact-QED limit (when $m_H$ formally tends to infinity) $D_{m_H}^{(2)}(m_W^{-1})\to 0$,
and $\xi\to\beta\zeta$,
as it should be. In the general case, when the mass of the Higgs field is moderate
and does not exceed $m_W$, we obtain
$\xi\propto\exp\left[-\frac{4\pi}{g^2}\left(m_W\epsilon+T\ln\left(\frac{{\rm e}^{\gamma}}{2}c\right)
\right)\right]$.
Here, we have introduced the notation $c\equiv m_H/m_W$, $c<1$, and $\gamma\simeq 0.577$ is the 
Euler constant, so that $\frac{{\rm e}^{\gamma}}{2}\simeq 0.89<1$. We see that 
the modified fugacity remains exponentially small, provided that 

\begin{equation}
\label{3}
T<-\frac{m_W\epsilon}{\ln\left(\frac{{\rm e}^{\gamma}}{2}c\right)}.
\end{equation}

This constraint should be updated by another one, which would provide the convergence
of the cumulant expansion applied in course of the average over $\psi$. Were the cumulant expansion 
divergent, this fact would indicate that the Higgs vacuum loses its normal stochastic properties and 
becomes a coherent one.
In order to get the 
new constraint, notice that the parameter of the cumulant expansion reads 
$\xi I^{(2)}$, where $I^{(2)}\equiv\int d^2x{\mathcal K}^{(2)}({\bf x})$.
The most essential, exponential, part of the parameter
of the cumulant expansion then reads~\cite{plb}:
$\xi I^{(2)}\propto\exp\left[-\frac{4\pi}{g^2}\left(m_W\epsilon+T
\ln\left(\frac{{\rm e}^{\gamma}}{2}c\right)-T\frac{\sqrt{2\pi}}{{\rm e}}\right)\right]$.
Therefore, the cumulant expansion converges at the temperatures obeying the inequality

$$T<\frac{m_W\epsilon}{\frac{\sqrt{2\pi}}{{\rm e}}-\ln\left(\frac{{\rm e}^{\gamma}}{2}c\right)},$$
which updates the inequality~(\ref{3}). On the other hand, since we are working in the plasma phase, i.e., 
$T\le g^2/2\pi$, it is enough to impose the following upper 
bound on the parameter of the weak-coupling approximation, $g^2/m_W$:

$$\frac{g^2}{m_W}<\frac{2\pi\epsilon}{\frac{\sqrt{2\pi}}{{\rm e}}-
\ln\left(\frac{{\rm e}^{\gamma}}{2}c\right)}.$$
Note that although this inequality is satisfied automatically at $\frac{{\rm e}^{\gamma}}{2}c\sim 1$, since  
it then takes the form $\frac{g^2}{m_W}<\sqrt{2\pi}{\rm e}\epsilon$, this is not so for the 
Bogomolny-Prasad-Sommerfield limit,
$c\ll 1$. Indeed, in such a case,
we have $\frac{g^2}{m_W}\ln\left(\frac{2}{c{\rm e}^{\gamma}}\right)<2\pi\epsilon$, 
that owing to the logarithm is however quite feasible.

\section{Including massless quarks}
Let us consider the extension of the model~(\ref{GG}) by
the fundamental dynamical quarks, which are supposed to be massless:
$\Delta S=-i\int d^3x\bar\psi\vec\gamma\vec D\psi$.
In this formula, $D_\mu\psi=\left(\partial_\mu-ig\frac{\tau^a}{2}A_\mu^a
\right)\psi$, $\bar\psi=\psi^{\dag}\beta$, where 
the Euclidean Dirac matrices are defined as $\vec\gamma=
-i\beta\vec\alpha$ with
$\beta=\left(
\begin{array}{cc}
1& 0\\
0& -1
\end{array}
\right)$ and 
$\vec\alpha=\left(
\begin{array}{cc}
0& \vec\tau\\
\vec\tau& 0
\end{array}
\right)$. Our discussion in the present Section will further follow Ref.~\cite{plb1}.
In that paper, it has been shown that at the temperatures higher than the BKT one,
quark zero modes in the monopole field lead to the additional attraction
between a monopole and an antimonopole in the molecule. 
In particular, when the number of these modes (equal to the
number of massless flavors) is sufficiently
large, the molecule shrinks so strongly that its size becomes of the order
of the inverse W-boson mass. Another factor which determines the size of the
molecule is the characteristic range of localization of zero modes. Namely, it can be shown that
the stronger zero modes are localized in the vicinity of the monopole center, the
smaller molecular size is. Let us consider the case when 
the Yukawa coupling of quarks with the Higgs field vanishes,
and originally massless quarks do not acquire any mass. This means that
zero modes are maximally delocalized. We shall see that in the case of one flavor, such
a weakness of the quark-mediated interaction of monopoles
opens a possibility for molecules to
undergo the phase transition
into the plasma phase at the temperature comparable with the BKT one.

It is a well known fact that in 3D, 't Hooft-Polyakov monopole is actually
an instanton~\cite{1}. Owing to this fact, we can use the results of Ref.~\cite{6} on the quark
contribution to the effective action of the instanton-antiinstanton molecule
in QCD. Referring the reader for details to Ref.~\cite{plb1}, we shall present here the final expression 
for the effective action, which reads $\Gamma=2N_f\ln|a|$. Here, $a=\left<\psi_0^{\bar M}\left|g\vec\gamma\frac{\tau^a}{2}
\vec A^{a{\,}M}\right|\psi_0^M\right>$ is the matrix element of the monopole field $\vec A^{a{\,}M}$ taken between 
the zero modes $\Bigl|\psi_0^M\Bigr>$, $\Bigl|\psi_0^{\bar M}\Bigr>$
of the operator $-i\vec\gamma\vec D$ defined at the field of
a monopole and an antimonopole, respectively. The dependence of $|a|$ on the molecular size $R$ 
can be straightforwardly found and reads
$|a|\propto\int d^3r/\left(r^2\left|\vec r-\vec R\right|\right)\simeq
-4\pi\ln(\mu R)$, where $\mu$ stands for the IR cutoff. 

At finite temperature, in the 
dimensionally-reduced
theory, the usual Coulomb interaction of monopoles~\footnote{Without
the loss of generality, we consider the molecule with the temporal component
of $\vec R$ equal to zero.}
$R^{-1}=
\sum\limits_{n=-\infty}^{+\infty}\left({\mathcal R}^2+(\beta n)^2\right)^{-1/2}$ 
goes over to $-2T\ln(\mu{\mathcal R})$, where ${\mathcal R}$ denotes the
absolute value of the 2D vector $\vec{\mathcal R}$. 
As far as the novel logarithmic interaction,
$\ln(\mu R)= \sum\limits_{n=-\infty}^{+\infty}\ln\left[\mu
\left({\mathcal R}^2+(\beta n)^2\right)^{1/2}\right]$, is concerned, it
transforms into
$\pi T{\mathcal R}+\ln\left[1-\exp(-2\pi T{\mathcal R})\right]-\ln 2$.
Accounting for both 
interactions and introducing the notation
$I\equiv\left[\pi T{\mathcal R}+\ln\left[1-\exp(-2\pi T{\mathcal R})\right]-\ln 2\right]^{-2N_f}$,
we eventually arrive at the following expression for the
mean squared separation $\left<L^2\right>$ in the molecule as a function of 
$T$, $g$, and $N_f$:

$$
\left<L^2\right>=\frac{
\int\limits_{m_W^{-1}}^{\infty}d{\mathcal R}{\mathcal R}^{3-\frac{8\pi T}{g^2}}I}
{\int\limits_{m_W^{-1}}^{\infty}d{\mathcal R}{\mathcal R}^{1-\frac{8\pi T}{g^2}}I}.
$$
At large ${\mathcal R}$, $\ln 2~ \& \bigl|\ln\left[1-\exp(-2\pi T{\mathcal R})\right]\bigr|
\ll \pi T{\mathcal R}$.
Hence, we see that $\left<L^2\right>$ is finite at
$T>(2-N_f)g^2/ 4\pi$, that reproduces the standard result $g^2/2\pi$
at $N_f=0$. For $N_f=1$, the plasma phase
is still present at $T<g^2/ 4\pi$,
whereas for $N_f\ge 2$ the monopole ensemble exists only
in the molecular phase at any temperature larger than $\zeta^{1/3}$.
Clearly, at $N_f\gg \max\left\{1,{4\pi T}/{g^2}\right\}$,
$\sqrt{\left<L^2\right>}\to m_W^{-1}$,
i.e., such a large number of zero modes shrinks the molecule
to the minimal admissible size.

\section*{Acknowledgments}
The author is grateful to Dr. N.O.~Agasian for collaboration and to Prof. A.~Di~Giacomo for useful discussions.
This work has been supported by INFN and partially by the INTAS grant Open Call 2000, Project No. 110. 
And last but not least, the author is grateful to the organizers
of the International Conference ``QCD02''
(Montpellier, 2-9th July 2002)
for an opportunity to present these results in a very pleasant and stimulating 
atmosphere.

\end{document}